\documentclass[11pt,a4paper]{article}
\usepackage{hyperref}
\usepackage{amssymb,amsmath}
\usepackage{tikz}
\usepackage{xspace}

\renewcommand{\vec}[1]{\mathbf{#1}}

\newcommand{\smat}{\left(\begin{smallmatrix}}
\newcommand{\stam}{\end{smallmatrix}\right)}
\newcommand{\mat}{\left(\begin{matrix}}
\newcommand{\tam}{\end{matrix}\right)}
\newcommand{\idop}{\mathbb{I}}
\newcommand{\iMPS}{$\infty$\,MPS\xspace}

\newcommand{\cC}{\mathcal{C}}
\newcommand{\cH}{\mathcal{H}}  
\newcommand{\cP}{\mathcal{P}}
\newcommand{\cV}{\mathcal{V}}

\newcommand{\blue}{\color{blue}}
\newcommand{\red}{\color{red}}

\title{Conformal field theory and the non-abelian $SU(2)_k$ chiral spin liquid}

\author{Thomas Quella$^{1}$ and Abhishek Roy
\\[2mm]\\$^1$\,The University of Melbourne\\School of Mathematics and Statistics\\Parkville 3010 VIC, Australia
\\[5mm]{\sc Thomas.Quella@unimelb.edu.au\qquad aroy@thp.uni-koeln.de}}

\date{}

\begin{document}
\maketitle
\begin{abstract}
  We construct a family of 1D and 2D long-range $SU(2)$ spin models as parent Hamiltonians associated with infinite dimensional matrix product states that arise from simple current correlation functions in $SU(2)_k$ WZW models. Our results provide a conformal field theory foundation for recent proposals by Greiter and coauthors regarding the realization of non-abelian chiral spin liquids. We explain, in particular, how the symmetrization procedure for the amplitudes of the ground state wave function suggested in \cite{Greiter:2009PhRvL.102t7203G} originates from the conformal field theory description.
\end{abstract}

\section{Introduction}

  Topological states of matter in two dimensions have been the subject of intensive study since the discovery and theoretical explanation of the fractional quantum Hall (FQH) effect \cite{Tsui:1982PhRvL..48.1559T,Laughlin:1983PhRvL..50.1395L}. These states generally arise as the effect of strong interactions and correlations and are thus hard to address from first principles. Most of the theoretical work since then has therefore focused on the description of idealized trial wave functions for ground states and excited states that exhibit the desired physical features, such as topological degeneracies and abelian or non-abelian anyonic statistics \cite{Laughlin:1983PhRvL..50.1395L,Moore:1991ks}. Many of these arise in a very natural way from 2D conformal field theory (CFT) which, all at the same time, encodes wave functions, braiding statistics of anyonic excitations and the spectrum of gapless edge modes. These investigations were complemented by the study of so-called parent Hamiltonians involving suitable pseudo-potentials that render the trial wave functions exact eigenstates \cite{Haldane:1983PhRvL..51..605H}. By now there is an enormous amount of literature on trial wave functions for ground states and excited states in fractional quantum Hall systems.
  
  
  Similar wave functions have also been proposed for the description of quantum spin liquid states \cite{Kalmeyer:1987PhRvL..59.2095K,Arovas:1988PhRvL..60..531A,Kalmeyer:1989PhRvB..3911879K} in 2D spin systems making use of a mapping between spin configurations and hardcore bosons. While originally only abelian chiral spin liquids were discussed, in Ref.~\cite{Greiter:2009PhRvL.102t7203G,Greiter:2011jp,Scharfenberger:2011PhRvB..84n0404S} Greiter and coauthors suggested a non-abelian generalization for each non-trivial choice of $SU(2)$ spin~$S$. These studies were accompanied by the derivation of a suitable parent Hamiltonian that annihilates the ground state wave function of these non-abelian chiral spin liquids \cite{Schroeter:2007PhRvL..99i7202S,Thomale:2009PhRvB..80j4406T,Greiter:2011jp}.

  Recently, the construction of topological and other exotic states of matter was picked up again from the perspective of quantum information theory. Numerical studies showed that the bipartite entanglement spectrum of FQHE ground states contains information about the gapless edge modes and hence about the underlying conformal field theory \cite{Li:2008PhRvL.101a0504L}. This idea can be turned around by trying to encode desired entanglement features into ground state wave functions. This is achieved by coupling the physical degrees of freedom to each other through an auxiliary layer of virtual spins or particles that encodes the entanglement. On a technical level this naturally leads to the concept of tensor network states. In that context there is another natural definition of parent Hamiltonians as a frustration-free combination of local interactions that annihilate the desired ground state locally.
  
  In the context of 2D topological quantum states it seems appropriate to model the entanglement layer through infinite dimensional matrix product states based on conformal field theory (CFT) correlators since many trial wave functions arise directly from CFT~\cite{Moore:1991ks,Read:1999PhRvB..59.8084R}. This program has been initiated in Ref.~\cite{Cirac:2010PhRvB..81j4431C} for free bosons and the $SU(2)_1$ WZW model and then further elaborated on in Ref.~\cite{Nielsen:2011py} for the $SU(2)_k$ WZW models, with strong emphasis on $k=1$ and $k=2$. While this, obviously, produced the same quantum states that were discussed previously based on the same CFTs, the approach gave natural access to (long range) parent Hamiltonians by employing identities for correlation functions that are associated with so-called null vectors, hence with the representation theory of the underlying affine Lie algebra symmetry. The same kind of construction was also employed for $SO(N)$ and $SU(N)$ spin models associated with $SO(N)_1$ and $SU(N)_1$ WZW models \cite{Tu:2013PhRvB..87d1103T,Tu:2014NuPhB.886..328T,Bondesan:2014NuPhB.886..483B}.
  
  Remarkably, in all $SU(N)$ cases it was shown that the parent Hamiltonians for $k=1$ readily reduces to a (trivial) variation of the long-range Haldane-Shastry model \cite{Haldane:1988PhRvL..60..635H,Shastry:1988PhRvL..60..639S} if the field insertions defining the \iMPS are chosen to lie equidistantly on the unit circle.
  This is exciting since the Haldane-Shastry model is a paradigmatic model for excitations with purely statistical interacions \cite{Haldane:1991PhRvL..67..937H}, is exactly solvable due to an exact Yangian symmetry \cite{Haldane:1992PhRvL..69.2021H} and is known to provide an accurate realization of the $SU(N)_1$ WZW models in the thermodynamic limit. 

  In a somewhat disconnected line of development, attempts were made to identify Hamiltonians that realize $SU(2)$ quantum spin systems in one dimension with higher order critical behaviour that is described by an $SU(2)_k$ Wess-Zumino-Witten (WZW) CFT \cite{Gepner:1986wi}. While these CFTs are known to arise from integrable systems based on $R$-matrices associated with the spin $\frac{k}{2}$-representation of $SU(2)$ \cite{Babujian:1983NuPhB.215..317B,Reshetikhin:1991JPhA...24.3299R}, the associated Hamiltonians involve higher order spin-spin couplings and are extremely fine-tuned. Several authors have suggested simpler Hamiltonians that are still claimed to be lattice approximations of $SU(2)_k$ WZW models. The authors of Ref.~\cite{Thomale:2012PhRvB..85s5149T} proposed a family of long-range Hamiltonians, for arbitrary level $k$, with quartic three-spin interactions which are based on a 1D reduction of parent Hamiltonians of the 2D non-abelian chiral spin liquid. Shortly after, a truncated short ranged version was studied in \cite{Michaud:2012PhRvL.108l7202M,Michaud:2013PhRvB..87n0404M} for $k=2,3,4$ with numerically determined couplings. In both cases, the flow to the $SU(2)_k$ WZW model in the thermodynamic limit was supported by numerical studies.
  
  It is probably fair to say that there is currently no model for higher spin $S$ (or higher level $k$) that has the same simplicity, symmetry and hence theoretical appeal as the Haldane-Shastry model for $S=\frac{1}{2}$ ($k=1$).\footnote{The existence of simple Haldane-Shastry-like spin chains with Yangian symmetry has, in fact, been ruled out in \cite{Haldane:1994cond.mat..1001H} for $S>\frac{1}{2}$.}
  Moreover, it should be noted that the \iMPS construction based on the $SU(2)_k$ WZW model for $k=1$ and $k=2$ that was discussed in the literature \cite{Cirac:2010PhRvB..81j4431C,Nielsen:2011py} is based on free field theories, namely a single free boson or three free fermions respectively for which correlation functions are readily available. In contrast, the $SU(2)_k$ WZW models for $k>2$ are genuine interacting conformal field theories. It is thus worth to provide a systematic investigation of the \iMPS construction for higher values of $k$. This is also interesting in the context of the recent paper \cite{Greiter:2019arXiv190509728G} which studies the associated non-abelian statistics of such one-dimensional models.

  In this paper we study an infinite dimensional matrix product state ($\infty$\,MPS) based on the $SU(2)_k$ WZW model and present explicit formulas for long-range parent Hamiltonians based on null vector conditions and WZW Ward identities. This extends earlier considerations in \cite{Nielsen:2011py} to general level~$k$ and clarifies the CFT origin of the wave function for the non-abelian chiral spin liquid (as discussed in \cite{Greiter:2014PhRvB..89p5125G}). Let us briefly outline the structure of this paper.
 
  Section~\ref{sc:Setup} is devoted to a discussion of the physical setup for an \iMPS construction of a spin-$S$ chain and the presentation of arguments that constrain the associated CFT to be an $SU(2)_k$ WZW model with $k=2S$. In Section~\ref{sc:ParentHamiltonian} the null vectors in that CFT are then used to derive a family of operators that annihilate the \iMPS and hence allow the construction of a parent Hamiltonian. The precise form of the \iMPS is discussed in Section~\ref{sc:GroundState} and compared to the construction of the non-abelian chiral spin liquid as suggested in Ref.~\cite{Greiter:2009PhRvL.102t7203G}. The general parent Hamiltonian for the 2D setup is simplified in Section~\ref{sc:Circle} for equidistant positions on the circle to investigate the reduction to a 1D setup.
  
  Two Appendices~\ref{ap:Notation} and~\ref{ap:SchwingerBosons} are used to summarize some elementary identities for the Lie algebra $su(2)$ and provide a brief review of the Schwinger boson construction. The final Appendix~\ref{ap:GeneralG} addresses the form of simple currents in $G_k$ WZW models in relation to those of products of $G_1$ WZW models for arbitrary (simple) symmetry groups $G$.

\section{\label{sc:Setup}Description of the physical setup}

  We wish to employ the idea of \iMPS and parent Hamiltonians \cite{Cirac:2010PhRvB..81j4431C,Nielsen:2011py} to construct a spin system consisting of $L$ spins $\vec{S}_l$ distributed arbitrarily on the complex plane. We will focus on $SU(2)$ spins and individual spins are supposed to transform in the spin-$S$ representation $\cV$ where $S=\frac{1}{2},1,\frac{3}{2},2,\ldots$ is fixed once and for all. The total Hilbert space of the system is thus $\cH=\cV^{\otimes L}$.
  To describe such a spin system within the formalism of \iMPS we, first of all, define a state
\begin{align}
  \label{eq:iMPS}
  |\psi\rangle
  =\sum_{m_1,\ldots,m_L}\bigl\langle\psi_{m_1}(z_1)\cdots\psi_{m_L}(z_L)\bigr\rangle_{\text{CFT}}\,|m_1\cdots m_L\rangle\;,
\end{align}
  where $|m_l\rangle$ denotes an orthonormal basis of $\cV$ and $\psi(z)$ are suitable primary fields in a 2D CFT that also transform in the representation $\cV$ and $\langle\cdots\rangle_{\text{CFT}}$ is the associated conformal block.\footnote{In standard CFT conventions, a primary field frequently is defined to be a field associated with a highest weight state. In contrast, in this paper a convention will be used where ``primary field'' refers to the whole ground state multiplet. In other words, our primary fields are $\cV$-valued. Here and in what follows we will, moreover, frequently suppress the index $m$ on $\psi$ to avoid cluttering of notation.} In addition, there are certain consistency requirements that will be addressed below. From the perspective of the state $|\psi\rangle$ the variables $z_l$ are merely parameters but we will interpret them as the positions of the quantum spins $\vec{S}_l$.
  
  In the case at hand, the natural CFTs are $SU(2)_k$ WZW models (with $k=1,2,\ldots$) since these come with the correct symmetry and primary fields. For such a CFT to exhibit a primary field transforming in the spin-$S$ representation we need to have $k\geq2S$. It should be noted that the fusion of primary fields may exhibit multiple channels and hence there is generally a whole space of conformal blocks $\bigl\langle\psi(z_1)\cdots\psi(z_L)\bigr\rangle_{\text{CFT}}$ whose dimension grows exponentially with the number of spins $L$. Hence the definition \eqref{eq:iMPS} is, in general, highly ambiguous and moving around the positions of the spins may lead to other states due to monodromies of conformal blocks~\cite{FrancescoCFT}.
  
   One (and in our situation the only) way to come up with a unique conformal block\footnote{Unique up to an overall phase which is irrelevant for the physical properties of the quantum state \eqref{eq:iMPS}.} is to choose the fields $\psi(z)$ to be simple currents. A simple current $\psi$ has, by definition, only a single fusion channel when fused with any other field $X$: $\psi\otimes_F X=Y$. In particular, for $SU(2)_k$ the fields are all self-dual and hence simple currents all fuse to the identity field: $\psi\otimes_F\psi=\idop$. This property immediately implies that the associated space of conformal blocks has dimension one (for even $L$) or zero (for odd $L$). We thus restrict our attention to even $L$ in what follows. The only non-trivial simple current of the $SU(2)_k$ WZW model transforms in the repesentation $S=\frac{k}{2}$. We are hence forced to assume $k=2S$ and this identification will be understood from now on.
   
  In Section~\ref{sc:ParentHamiltonian} we will define (or rather derive) operators $\cC_l(\{z_i\})$ on $\cH$ that take the positions $z_i$ as parameters and annihilate the quantum state $|\psi\rangle$. The associated parent Hamiltonian
\begin{align}
  \label{eq:ParentHamiltonianGeneral}
  H=\sum_{l=1}^L\cC_l\bigl(\{z_i\}\bigr)^\dag\cdot\cC_l\bigl(\{z_i\}\bigr)
\end{align}
  is then positive semi-definite (satisfies $H\geq0$) and $|\psi\rangle$ is a zero energy ground state of $H$. We will see that the Hamiltonian $H$ as defined in \eqref{eq:ParentHamiltonianGeneral} is closey related to the Hamiltonian found earlier in \cite{Greiter:2011jp} (without noting the relation to CFT) and further discussed in \cite{Thomale:2012PhRvB..85s5149T}.
  For $S=\frac{1}{2}$ and $S=1$ our results readily reduce to those presented in \cite{Nielsen:2011py}.

\section{\label{sc:ParentHamiltonian}Derivation of the parent Hamiltonian from CFT}

  We will construct the operators $\cC_l\bigl(\{z_i\}\bigr)$ based on our knowledge of null fields related to the simple currents $\psi(z)$ in the $SU(2)_k$ WZW model where $k=2S$. This procedure was suggested in \cite{Cirac:2010PhRvB..81j4431C,Nielsen:2011py} and carried out explicitly for $S=\frac{1}{2}$ and $S=1$ based on specific formulas for Clebsch-Gordan coefficients. In contrast, we will employ a formalism that is manifestly basis-independent and straightforwardly generalizes to higher values of $S$.
  
\begin{figure}
\begin{center}
\begin{tikzpicture}[line width=1pt]
  \draw[red!10!white,fill=red!10!white,rounded corners] (-2.2,-.2) rectangle (2.2,.2);
  \draw[red!10!white,fill=red!10!white,rounded corners] (-2.2,.7) rectangle (2.2,1.3);
  \draw[blue!10!white,fill=blue!10!white,rounded corners] (-3.2,.8) rectangle (3.2,1.2);
  \draw[-latex] (0,-.5) -- (0,2) node[left] {$L_0$};
  \draw[-latex] (-4.5,0) node[left] {$h_{S}$} -- (4.5,0) node[below] {$J_0^z$};
  \draw (-4,1) node[left] {$h_{S}+1$} -- (4,1);
  \foreach \x in {-1,0,1} {
    \draw[red] (\x,1) circle (6pt);
  };
  \foreach \x in {-2,-1,0,1,2} {
    \draw[red,fill] (\x,0) circle (2pt);
    \draw[red] (\x,1) circle (4pt);
  };
  \foreach \x in {-3,-2,-1,0,1,2,3} {
    \draw[blue,fill] (\x,1) circle (2pt);
  };
  \draw (2,0) node[below=1mm] {$S$};
  \draw (-2,0) node[below=1mm] {$-S$};
  \draw (3,0) +(0,.1) -- +(0,-.1);
  \draw (-3,0) +(0,.1) -- +(0,-.1);
  \draw (3,0) node[below=1mm] {$S{+}1$};
  \draw (-3,0) node[below=1mm] {$-S{-}1$};
  \draw[-latex,shorten <=2mm,shorten >=2mm] (2,0) -- node[right=1mm] {$J_{-1}^+$} (3,1);
  \draw[-latex,shorten <=2mm,shorten >=2mm] (2,0) -- (2,1);
  \draw[-latex,shorten <=2mm,shorten >=2mm] (2,0) -- node[left=1mm] {$J_{-1}^-$} (1,1);
  \draw[-latex,shorten <=2mm,shorten >=2mm] (-2,0) -- node[left=1mm] {$J_{-1}^-$} (-3,1);
  \draw[-latex,shorten <=2mm,shorten >=2mm] (-1,0) -- node[left] {$J_{-1}^z$} (-1,1);
   \draw (1,1.7) node {$\vdots$};
   \draw (2,1.7) node {$\vdots$};
   \draw (3,1.7) node {$\vdots$};
  \draw (-1,1.7) node {$\vdots$};
  \draw (-2,1.7) node {$\vdots$};
  \draw (-3,1.7) node {$\vdots$};
  \draw (8,1) node {${\blue(S+1)}\oplus{\red(S)}\oplus{\red(S-1)}$};
  \draw (8,0) node {\red$(S)$};
  \draw[-latex,shorten <=3mm,shorten >=2mm] (8,0) -- node[right] {$\otimes1$} (8,1);
  \draw (8,1.7) node {Representation content w.r.t.\ $SU(2)$};
\end{tikzpicture}
  \caption{\label{fig:ModuleS} (Color online) Low lying states in the $SU(2)_k$ Verma module  associated with the spin-$S$ representation of $SU(2)$ (where $k=2S$). The states belonging to the representation~$(S+1)$ on energy level~$1$ (blue dots) are null states. The circles indicate non-trivial degeneracies.}
\end{center}
\end{figure}
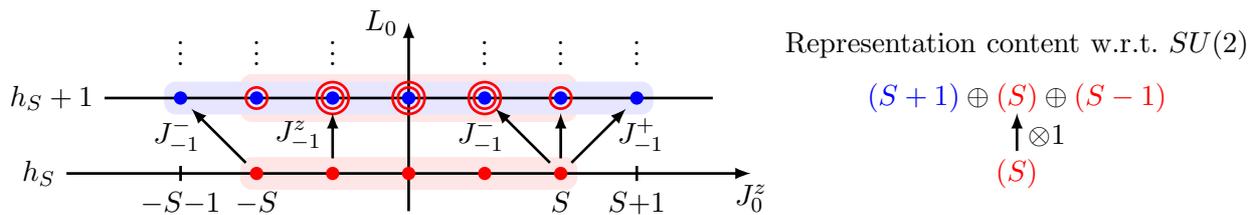

  Since the primary field $\psi(z)$ is transforming in the representation $S$, the associated null vector is located at energy level 1, regardless of the value of $S$ (see, e.g., \cite{FrancescoCFT}). The states on energy level 1 can all be represented as $J_{-1}^a|m\rangle$ where $m=-S,-S+1,\ldots,S$ is the magnetic quantum number and $J_n^a$ are the modes of the affine Lie algebra $\widehat{su}(2)_k$ underlying the $SU(2)_k$ WZW model. These states form a representation with respect to the global symmetry group $SU(2)$ which is generated by the zero modes $J_0^a$ and decomposes as
\begin{align}
  \label{eq:TensorProduct}
  1\otimes S
  =(S+1)\oplus(S)\oplus(S-1)\;,
\end{align}
  where $1$ arises since the modes $J_{-1}^a$ transform in the adjoint representation and $S$ refers to the action on the ground states $|m\rangle$.\footnote{Mathematically, this follows from $J_0^a(J_{-1}^b\otimes|m\rangle)=[J_0^a,J_{-1}^b]\otimes|m\rangle+J_{-1}^b\otimes J_0^a|m\rangle$ where we introduced (superfluous) tensor product symbols $\otimes$ to make the connection more explicit.} An illustration of these statements can be found in Figure~\ref{fig:ModuleS}.
  
  The null vectors lie in the top component $(S+1)$ of the tensor product decomposition \eqref{eq:TensorProduct}. Let $\vec{t}$ and $\vec{S}$ be the spin generators associated with the action of $SU(2)$ on $J_{-1}^a$ and $|m\rangle$, respectively. Then $C=(\vec{t}+\vec{S})^2$ denotes the quadratic Casimir operator on this tensor product and the projector onto the null states is given by
\begin{align}
  \cP
  =\frac{(C-C_S)(C-C_{S-1})}{(C_{S+1}-C_S)(C_{S+1}-C_{S-1})}\;,
\end{align}
  where we have made use of the Casimir eigenvalues $C_J=J(J+1)$.
  Upon expanding this product, one finds the projector
\begin{align}
  \cP
  =\frac{1}{2S+1}+\frac{S+2}{(S+1)(2S+1)}\,\vec{t}\cdot\vec{S}
       +\frac{(\vec{t}\cdot\vec{S})^2}{(S+1)(2S+1)}\;.
\end{align}
  This expression may be further simplified since $\vec{t}$ is a spin in the adjoint representation and hence specified by matrix elements
  ${(t^e)^a}_c=-i{\epsilon^{ea}}_c$ where $\epsilon$ is the completely anti-symmetric Levi-Civita symbol. Using identities summarized in Appendix~\ref{ap:Notation} we then find
\begin{align}
  {\bigl[\vec{t}\cdot\vec{S}\bigr]^{ab}}_{cd}
  =-i{\epsilon^a}_{ce}\,{(S^e)^b}_d
  \quad\text{ and }\quad
  {\bigl[(\vec{t}\cdot\vec{S})^2\bigr]^{ab}}_{cd}
  ={\bigl[S(S+1)\,\delta_c^a\,\idop-S_cS^a\bigr]^b}_d\;,
\end{align}
  where we have emphasized the operator structure in the second tensor factor while keeping explicit matrix indices for the first one. Writing the projector as a matrix in the first two indices only (i.e.\ as an
  operator valued matrix), one then obtains
\begin{align}
  {\cP^a}_b
  &=\delta_b^a\,\frac{S+1}{2S+1}\,\idop
        -\frac{i(S+2)}{(S+1)(2S+1)}\,{\epsilon^a}_{bc}\,S^c
        -\frac{S_bS^a}{(S+1)(2S+1)}\;.
\end{align}
  This expression can be symmetrized using the commutation relations of $SU(2)$ and this leads to the final result
\begin{align}
  \label{eq:ProjectorSymmetrized}
  {\cP^a}_b
  &=\delta_b^a\,\frac{S+1}{2S+1}\,\idop
        -\frac{i(2S+3){\epsilon^a}_{bc}\,S^c}{2(S+1)(2S+1)}
        -\frac{S^aS_b+S_bS^a}{2(S+1)(2S+1)}\;.
\end{align}
  The symmetrization will facilitate a comparison of our parent Hamiltonian \eqref{eq:ParentHamiltonianGeneral} with that proposed in Ref.~\cite{Greiter:2011jp}.

  By way of construction of the projector ${\cP^a}_b$, we are guaranteed that ${\cP^a}_b\psi(z)$ is a null field.\footnote{We suppress the indices in $\psi(z)$ in order to avoid cluttering of notation.} Consequently, all correlation functions involving this field vanish and this allows to conclude, after using a WZW Ward identity, that
\begin{align}
  \label{eq:NullFieldCorrelator}
  0={\cP_l^a}_b\,\bigl\langle\psi(z_1)\cdots
         J_{-1}^b\psi(z_l)\cdots\psi(z_L)\bigr\rangle
   =\cP_l^a\bigl(\{z_l\}\bigr)\,\bigl\langle\psi(z_1)\cdots\psi(z_L)\bigr\rangle\;,
\end{align}
  where the operator appearing on the right hand side is defined by\footnote{Here and in what follows we write $j(\neq l)$ to emphasize that the summation is over $j$ while $l$ is fixed. In contrast, $j\neq l$ would mean a sum over $j$ and $l$.}
\begin{align}
  \label{eq:OperatorP}
  \cP_l^a\bigl(\{z_i\}\bigr)
  =\sum_{j(\neq l)}\frac{{\cP_l^a}_b\,S_j^b}{z_l-z_j}\;.
\end{align}
  The subscript $l$ on ${\cP_l^a}_b$ indicates that the spin operators $\vec{S}$ present in this matrix, see Eq.~\eqref{eq:ProjectorSymmetrized}, act on the $l$'s tensor factor and hence can be written $\vec{S}_l$. By Eq.~\eqref{eq:NullFieldCorrelator} all operators $\cP_l^a\bigl(\{z_i\}\bigr)$ annihilate the \iMPS defined in Eq.\ \eqref{eq:iMPS},
\begin{align}
  \label{eq:PAction}
  \cP_l^a\bigl(\{z_i\}\bigr)|\psi\rangle
  =\sum_{j(\neq l)}\frac{{\cP_l^a}_b\,S_j^b}{z_l-z_j}\,|\psi\rangle
  =0\;,
\end{align}
  and can be used to define an associated parent Hamiltonian along the lines of Eq.~\eqref{eq:ParentHamiltonianGeneral}.
  
  Before we spell out the parent Hamiltonian we will slightly generalize the coordinate dependence of \eqref{eq:OperatorP}. This is possible in view of the equations
\begin{align}
  \label{eq:ModPAction}
  \sum_jS_j^b|\psi\rangle=0\;,\qquad
  {\cP_l^a}_bS_l^b
  &=0\;
  \quad\text{ and hence }\quad
  0=\sum_{j(\neq l)}{\cP_l^a}_b S_j^b|\psi\rangle\;.
\end{align}
  The first equation is just the statement that $|\psi\rangle$ is a singlet, the second equation follows from a straightforward calculation and the last equality is a simple consequence of the former two. We can then combine Eqs.~\eqref{eq:PAction} and \eqref{eq:ModPAction} with arbitrary coefficients to obtain
\begin{align}
  \label{eq:OperatorC}
  \cC_l^a\bigl(\{z_i\}\bigr)|\psi\rangle
  =0
  \quad\text{ where }\quad
  \cC_l^a\bigl(\{z_i\}\bigr)
  =\sum_{j(\neq l)}\Omega_{lj}\,{\cP_l^a}_b\,S_j^b
\end{align}
  for relatively general choices of the parameters $\Omega_{lj}$ (as functions of the $z_i$). A few of these choices will be discussed below in Section~\ref{sc:Circle}.

  The operators $\cC_l^a\bigl(\{z_i\}\bigr)$ are used to construct the parent Hamiltonian defined in Eq.~\eqref{eq:ParentHamiltonianGeneral}. The evaluation of this expression is greatly simplified by the fact that the matrices ${\cP_l^a}_b$ are projectors and hermitean. The resulting terms are all straightforward to bring into a convenient form except for the last one which is quartic in spin operators. Upon using the commutation relations and some other identities listed in Appendix~\ref{ap:Notation} we then ultimately find
\begin{align}
  \label{eq:HSymmetrized}
  H&=\sum_{l=1}^L\cC_l^a\bigl(\{z_i\}\bigr)^\dag\kappa_{ab}\,\cC_l^b\bigl(\{z_i\}\bigr)
    =\sum_k\sum_{i,j(\neq k)}\bar{\Omega}_{ki}\Omega_{kj}\biggl[
        \frac{S+1}{2S+1}\,\vec{S}_i\cdot\vec{S}_j
        -\frac{\delta_{ij}\,\vec{S}_k\cdot\vec{S}_i}{2(S+1)(2S+1)}\nonumber\\[2mm]
  &\qquad\qquad\qquad+\frac{i(2S+3)}{2(S+1)(2S+1)}
    \,(\vec{S}_i\times\vec{S}_k)\cdot\vec{S}_j
   -\frac{(\vec{S}_i\cdot\vec{S}_k)(\vec{S}_k\cdot\vec{S}_j)
    +(\vec{S}_j\cdot\vec{S}_k)(\vec{S}_k\cdot\vec{S}_i)}{2(S+1)(2S+1)}\biggr]\;.
\end{align}
  We note that this Hamiltonian features bilinear and biquadratic interactions\footnote{Biquadratic interactions arise from the quartic interactions when choosing $i=j$.} as well as three- and four-spin terms. The three-spin term is special in the sense that it is chiral, i.e.\ not invariant under time-reversal or a reordering of the spins. However, it may well be absent for specific choices of the parameters $\Omega_{ij}$ as will be discussed in Section~\ref{sc:Circle}.

\section{\label{sc:GroundState}Conformal field theory description of the ground state}

  To determine the explicit form of the \iMPS ground state \eqref{eq:iMPS} we need to calculate the WZW correlation function
\begin{align}
  \label{eq:ConformalBlock}
  \bigl\langle\psi_{m_1}(z_1)\cdots\psi_{m_L}(z_L)\bigr\rangle\;.
\end{align}
  We will now show that this correlator follows from a symmetrization argument that closely resembles that given in \cite{Greiter:2011jp}, but here from the perspective of the underlying CFT.

  We first of all note that there is a diagonal embedding
\begin{align}
  \label{eq:EmbeddingNaive}
  SU(2)_k\subset\underbrace{SU(2)_1\times\cdots\times SU(2)_1}_{k\text{ factors}}\;.
\end{align}
  However, this embedding is not conformal since the central charges on the two sides of the equation do not coincide:
\begin{align}
  \frac{3k}{k+2}\neq k\times 1\qquad(\text{except for }k=1)\;.
\end{align}
  As we will show momentarily there is still a way of thinking of the simple current $\psi^{(k)}(z)$ in the $SU(2)_k$ WZW model as being built from the simple currents $\psi_\alpha^{(1)}(z)$ (with $\alpha=1,\ldots,k$) of the $SU(2)_1$ WZW models appearing on the right hand side of Eq.~\eqref{eq:EmbeddingNaive}.
  
 In a first step we make the embedding conformal by considering
\begin{align}
  \label{eq:Embedding}
  \frac{SU(2)_1\times\cdots\times SU(2)_1}{SU(2)_k}\times SU(2)_k
  \ \subset\ \underbrace{SU(2)_1\times\cdots\times SU(2)_1}_{k\text{ factors}}
\end{align}
  instead of Eq.~\eqref{eq:EmbeddingNaive}. The primary fields of the product theory are associated with representations $\cH_{(j_1,\ldots,j_L)}=\cH_{j_1}^{(1)}\otimes\cdots\otimes\cH_{j_L}^{(1)}$ where $j_i=0,\frac{1}{2}$. According to the usual branching rules, these decompose into a direct sum
\begin{align}
  \label{eq:RepDecomposition}
  \cH_{(j_1,\ldots,j_L)}
  =\bigoplus_{j=0}^{k/2}\,\cH_{(j_1,\ldots,j_L|j)}\otimes\cH_j^{(k)}
\end{align}
  under the embedding \eqref{eq:Embedding},
  where $\cH_{(j_1,\ldots,j_L|j)}$ is a representation of the diagonal coset  and $\cH_j^{(k)}$ one of $SU(2)_k$. Some of the coset representations are in fact trivial in the sense that they do not appear in the decomposition~\eqref{eq:RepDecomposition}. This is described by selection rules which, in the present case, have the form
\begin{align}
  \label{eq:SelectionRule}
  j_1+\cdots+j_L+j\equiv0\mod2\;.
\end{align}
  In other words, for a representation $j$ of $SU(2)_k$ to appear in the decomposition~\eqref{eq:RepDecomposition} it needs to satisfy Eq.~\eqref{eq:SelectionRule}. The simple current we are interested in corresponds to $j=\frac{k}{2}$ and we see that this arises very naturally from the product representation $\cH_{(\frac{1}{2},\ldots,\frac{1}{2})}$ which corresponds to a certain projection of the product of the $SU(2)_1$ simple currents in the product theory. On the other hand, the latter projection also gives rise to a coset field corresponding to $\cH_{(\frac{1}{2},\ldots,\frac{1}{2}|\frac{k}{2})}$. We hence still need to explain why this coset field does not play a role and, moreover, specify in more detail what projection to choose in order to single out the $j=\frac{k}{2}$ component of the composition~\eqref{eq:RepDecomposition}.
  
  Let us start with the former task. It is well-known that the presence of selection rules goes hand in hand with the presence of ``field identifications'' that in the present case read
\begin{align}
  \cH_{(j_1,\ldots,j_L|j)}
  \cong\cH_{(\frac{1}{2}-j_1,\ldots,\frac{1}{2}-j_L|\frac{k}{2}-j)}\;.
\end{align}
  We immediately recognize that this field identification implies
\begin{align}
  \cH_{(\frac{1}{2},\ldots,\frac{1}{2}|\frac{k}{2})}
  \cong\cH_{(0,\ldots,0|0)}\;,
\end{align}
  which is just the vacuum module of the diagonal coset theory. In other words, the simple current of the $SU(2)_k$ WZW model that we are after is paired with the identity field of the coset (or one of its descendants). A comparison of conformal dimensions (see below) implies that the remaining factor is indeed a multiple of the identity field and not a non-trivial descendant. This is good news since this means that the desired correlator of the $SU(2)_k$ simple current can be directly obtained from the product of correlators of the $SU(2)_1$ simple currents, without additional complications due to the coset part.
    
  From Young diagram techniques it is evident that the $\frac{k}{2}$ representation of $SU(2)$ arises as the top component in the tensor product $\frac{1}{2}\otimes\cdots\otimes\frac{1}{2}$ ($k$ factors) which is associated with the complete symmetrization of all factors. Hence we obtain the identity\footnote{The product is non-singular since the fields all live in different tensor factors.}
\begin{align}
  \label{eq:Psik}
  \psi^{(k)}(z)
  =\text{Symmetrization}\bigl(\psi_1^{(1)}(z)\cdots\psi_k^{(1)}(z)\bigr)
\end{align}
  relating the simple currents in $SU(2)_k$ and $k$ copies of $SU(2)_1$, where the symmetrization refers to the magnetic quantum numbers $m_\alpha=\pm\frac{1}{2}$ which have been supressed. We note that each of the fields $\psi_l^{(1)}(z)$ contributes $h^{(1)}=\frac{1}{4}$ to the total conformal dimension of the product since they mutually commute. This precisely matches $h^{(k)}=\frac{k}{4}$ for the field $\psi^{(k)}(z)$ on the left hand side\footnote{A spin $j$ primary field has conformal dimension $h_j=j(j+1)/(k+2)$ in the $SU(2)_k$ WZW model. The two results quoted are obtained for $j=\frac{k}{2}$, in the first case after restriction to $k=1$.} and indeed shows that relation \eqref{eq:Psik} holds without the insertion of a descendant in the coset part. For the determination of the desired conformal block~\eqref{eq:ConformalBlock} it thus suffices to have knowledge of the $SU(2)_1$ correlation functions which can be calculated by means of a free field construction~\cite{FrancescoCFT}. Instead of reproducing the derivation we just spell out the result \cite{Cirac:2010PhRvB..81j4431C},
\begin{align}
  \bigl\langle\psi_{m_1}^{(1)}(z_1)\cdots\psi_{m_L}^{(1)}(z_L)\bigr\rangle
  =\rho_{\vec{m}}\prod_{i<j}(z_i-z_j)^{2m_im_j}\;.
\end{align}
  The constant $\rho_{\vec{m}}$ appearing here is given by
\begin{align}
  \rho_{\vec{m}}
  =\prod_{i\text{ odd}}\rho_{m_i}
  \qquad\text{ where }\quad
  \rho_m=\begin{cases}
           \phantom{-}1&,\;m=\frac{1}{2}\\
           -1&,\;m=-\frac{1}{2}
         \end{cases}
\end{align}
  and is known as the Marshall sign factor.
  
  As was suggested in~\cite{Arovas:1988PhRvL..60..531A} in the context of AKLT models and then adapted to the non-abelian chiral spin liquid in~\cite{Greiter:2009PhRvL.102t7203G}, the symmetrization can be carried out quite explicitly in terms of Schwinger bosons, see Appendix~\ref{ap:SchwingerBosons}. In that formalism and setting $M=\frac{L}{2}$, the state that is relevant for the case $k=1$ is\footnote{The subscript $0$ indicates that this is meant to be a ground state.}
\begin{align}
  |\psi_0^{\text{KL}}\rangle
  =\sum_{\{\xi_1,\ldots,\xi_M\}}
   \psi_0^{\text{KL}}(z_{\xi_1},\ldots,z_{\xi_M})\,
   a_{\xi_1}^\dag\cdots a_{\xi_M}^\dag
   b_{\eta_1}^\dag\cdots b_{\eta_M}^\dag|0\rangle
  =\psi_0^{\text{KL}}[\vec{a}^\dag,\vec{b}^\dag]|0\rangle\;,
\end{align}
  where the sum extends over all subsets $\{\xi_1<\ldots<\xi_M\}\subset\{1,\ldots,L\}$, the subset $\{\eta_1<\ldots<\eta_M\}$ complements the previous one and all entries are ordered as indicated.
  This state may then easily be generalized to the spin-$S$ representation by defining~\cite{Greiter:2009PhRvL.102t7203G}
\begin{align}
  \label{eq:SpinS}
  |\psi_0^S\rangle
  =\bigl(\psi_0^{\text{KL}}[\vec{a}^\dag,\vec{b}^\dag]\bigr)^{2S}|0\rangle\;.
\end{align}
  The mathematical situation is depicted in the following diagram,
\begin{center}
\begin{tikzpicture}
  \foreach \y in {0,-1.2} {
    \draw (0,\y) node  {$\frac{1}{2}$};
    \draw (.3,\y) node  {$\otimes$};
    \draw (.6,\y) node  {$\frac{1}{2}$};
    \draw (.9,\y) node  {$\otimes$};
    \draw (1.2,\y) node  {$\frac{1}{2}$};
    \draw (1.5,\y) node  {$\otimes$};
    \draw (2,\y) node  {$\cdots$};
    \draw (2.5,\y) node  {$\otimes$};
    \draw (2.8,\y) node  {$\frac{1}{2}$};
    \draw (3.5,\y) node {$\rightarrow0$};
  };
  \draw (0,-.5) node {$\vdots$};
  \draw (.6,-.5) node {$\vdots$};
  \draw (1.2,-.5) node {$\vdots$};
  \draw (2.8,-.5) node {$\vdots$};
  \draw (.6,-.6) node[left] {${\footnotesize k\text{ copies }}\left\{\phantom{\begin{array}{c}X\\[1.2cm]\end{array}}\right.$};
  \draw (1.4,0) node[above] {$\overbrace{\phantom{XXXXXXXX}}^{L \text{ copies}}$};
  \draw (0,-1.8) node {$\downarrow$};
  \draw (.6,-1.8) node {$\downarrow$};
  \draw (1.2,-1.8) node {$\downarrow$};
  \draw (2.8,-1.8) node {$\downarrow$};
  \draw (0,-2.4) node  {$\frac{k}{2}$};
  \draw (.3,-2.4) node  {$\otimes$};
  \draw (.6,-2.4) node  {$\frac{k}{2}$};
  \draw (.9,-2.4) node  {$\otimes$};
  \draw (1.2,-2.4) node  {$\frac{k}{2}$};
  \draw (1.5,-2.4) node  {$\otimes$};
  \draw (2,-2.4) node  {$\cdots$};
  \draw (2.5,-2.4) node  {$\otimes$};
  \draw (2.8,-2.4) node  {$\frac{k}{2}$};
\end{tikzpicture}
\end{center}
 On the one hand we have the Hilbert space of $k$ individual spin systems (it is useful to think about them as chains here) which are each projected into a specific singlet by means of the \iMPS construction for $k=1$. On the other hand we have projections to the maximal spin component $\frac{k}{2}$ across the various spin systems for each individual site.\footnote{``Individual site'' here refers to an arbitrary numbering of spins which is identical on each of the $k$ spin systems.}

  From the perspective of the previous comments it is sensible to first define the state\footnote{We would like to emphasize that the subscript of the vectors $\vec{a}_i$ and $\vec{b}_i$ refers to different chains, not the sites on individual chains.}
\begin{align}
  |\psi_0\rangle_{k\text{ copies}}
  =\bigl(\psi_0^{\text{KL}}[\vec{a}_1^\dag,\vec{b}_1^\dag]\bigr)\cdots\bigl(\psi_0^{\text{KL}}[\vec{a}_k^\dag,\vec{b}_k^\dag]\bigr)|0\rangle\;.
\end{align}
  This is a singlet but it still lives in the Hilbert space $((\frac{1}{2})^{\otimes L})^{\otimes k}=(\frac{1}{2})^{\otimes kL}$. The desired projection onto $(\frac{k}{2})^{\otimes L}$ is implemented by means of a complete symmetrization, i.e.\ by identifying all the bosons: $(\vec{a}_i,\vec{b}_i)=(\vec{a},\vec{b})$. This procedure leads directly to the quantum state \eqref{eq:SpinS}. This argument proves that the \iMPS underlying our construction of the parent Hamiltonian \eqref{eq:HSymmetrized} indeed agrees with the state \eqref{eq:SpinS} introduced by Greiter and Thomale \cite{Greiter:2009PhRvL.102t7203G}.
  
\section{\label{sc:Circle}Equidistant spins on the unit circle}

  The Hamiltonian \eqref{eq:HSymmetrized} can be simplified significantly if the parameters $z_l$ are assumed to be distributed equidistantly on the unit circle, i.e.\ $z_l=e^{\frac{2\pi i}{L}l}$, and the functions $\Omega_{ij}$ are chosen appropriately.\footnote{Many of the simplifications we present carry through even for arbitrary positions $z_l=e^{i\phi_l}$ on the unit circle where the $z_l$ are assumed to be mutually distinct. Since our main motivation is the comparison of the \iMPS approach with the recent proposal of Greiter and coauthors \cite{Greiter:2011jp} we refrain from presenting the general expressions.} Natural choices for $\Omega_{ij}$ that are consistent with the derivation of the operators $\cC_l$ in Eq.~\eqref{eq:OperatorC} are
\begin{align}
  x_{ij}=\frac{1}{z_i-z_j}\;,\quad
  w_{ij}=\frac{z_i+z_j}{z_i-z_j}
  \quad\text{ and }\quad
  \theta_{ij}=\frac{z_i}{z_i-z_j}\;.
\end{align}  
  On the unit circle one has $\bar{z}_i=1/z_i$ and this allows to simplify expressions such as $|\Omega_{ij}|^2$ or $\bar{\Omega}_{li}\Omega_{lj}$ that appear in the \iMPS parent Hamiltonian.

  In what follows we restrict our attention to the choice $w_{ij}=\frac{z_i+z_j}{z_i-z_j}$. The most significant benefit of this choice is the absence of the cross product term $(\vec{S}_i\times\vec{S}_k)\cdot\vec{S}_j$ in Eq.~\eqref{eq:HSymmetrized} whenever $i\neq j$. Indeed, this term cancels out in view of the relation $\bar{w}_{ij}=-w_{ij}$ and the resulting symmetry of $\bar{w}_{ki}w_{kj}=-w_{ki}w_{kj}$ in the indices $i$ and $j$. For $i=j\neq k$ on the other hand we can use the identities of Appendix~\ref{ap:Notation} to write
\begin{align}
  i(\vec{S}_i\times\vec{S}_k)\cdot\vec{S}_j
  =\vec{S}_i\cdot\vec{S}_k\;,
\end{align}
  which is bilinear. It is convenient to collect all bilinears into a single sum and to simplify all the constant contributions that arise on the way. To do so we first of all evaluate the sums \cite{Nielsen:2011py}
\begin{align}
  \label{eq:WGeneral}
  \sum_{k(\neq i,j)}\bar{w}_{ki}w_{kj}
  =2-L-2w_{ij}^2
  \quad\text{ and }\quad
  \sum_{k\neq l}|w_{kl}|^2
  =\frac{1}{3}L(L-1)(L-2)\;.
\end{align}
  After using the first expression in Eq.~\eqref{eq:HSymmetrized}, the Hamiltonian features spin-spin interactions without dependence on the parameters $z_l$. These can be rewritten in terms of the identity operator and the square $\vec{S}_{\text{tot}}^2$ of the total spin operator $\vec{S}_{\text{tot}}$ using the identity
$  \sum_{k\neq l}\vec{S}_k\cdot\vec{S}_l
  =\vec{S}_{\text{tot}}^2-LS(S+1)$.
  After collecting all terms of the same form we then arrive at\footnote{In writing down this expression we rescaled the original Hamiltonian~\eqref{eq:HSymmetrized} by a factor $\frac{2\pi^2(2S+1)}{4L^2(2S+3)}$. This is possible without changing any of the fundamental properties of $H$. The rescaling ensures that the Hamiltonian behaves as expected in the thermodynamic limit $L\to\infty$ and is required for a comparison with the results of~\cite{Greiter:2011jp}.}
\begin{align}
  \label{eq:HPreFinal}
  H&=\frac{2\pi^2}{L^2}\Biggl[\frac{LS(S+1)^2(L-2)(L+2)}{12(2S+3)}
  -\frac{1}{4}\sum_{i\neq j}w_{ij}^2
   \,\vec{S}_i\cdot\vec{S}_j\nonumber\\[2mm]
  &\qquad\qquad
   -\sum_k\sum_{i,j(\neq k)}\bar{w}_{ki}w_{kj}\frac{(\vec{S}_i\cdot\vec{S}_k)(\vec{S}_k\cdot\vec{S}_j)
    +(\vec{S}_j\cdot\vec{S}_k)(\vec{S}_k\cdot\vec{S}_i)}{8(S+1)(2S+3)}
    -\frac{(S+1)(L-2)}{4(2S+3)}\,\vec{S}_{\text{tot}}^2\Biggr]\;.
\end{align}
  We note that the four-spin term includes biquadratic spin couplings for $i=j$.
  
  It remains to obtain a better idea about the coordinate dependence of the interactions, i.e.\ the dependence on the parameters $z_l$. For the bilinear term the relevant expression is
\begin{align}
  \label{eq:wij}
  |w_{ij}|^2
  =-w_{ij}^2
  =-\frac{4z_iz_j}{(z_i-z_j)^2}-1
  =\frac{4}{|z_i-z_j|^2}-1
  =\frac{1}{\sin^2\frac{\pi}{L}(i-j)}-1\;.
\end{align}
  This clearly reproduces the distance dependence familiar from the Haldane-Shastry Hamiltonian.
  The simplification of the expression $\bar{w}_{ki}w_{kj}$ requires slightly more thoughts but eventually it is a matter of elementary algebra to show that
\begin{align}
  \bar{w}_{ki}w_{kj}
  =\frac{2}{(\bar{z}_k-\bar{z}_i)(z_k-z_j)}
   +\frac{2}{(\bar{z}_k-\bar{z}_j)(z_k-z_i)}-1\;.
\end{align}
  We note that the last expression is multiplying a term that is symmetric in the indices $i$ and $j$. Hence the summation simplifies and we are left with the final expression
\begin{align}
  \label{eq:HFinal}
  H&=\frac{2\pi^2}{L^2}\Biggl\{\frac{LS(S+1)\bigl(L^2(S+1)+2S+5\bigr)}{12(2S+3)}
     +\sum_{k\neq l}\frac{\vec{S}_k\cdot\vec{S}_l}{|z_k-z_l|^2}
     -\frac{L(S+1)+1}{4(2S+3)}\,\vec{S}_{\text{tot}}^2\nonumber\\[2mm]
  &\qquad\qquad
   -\sum_k\sum_{i,j(\neq k)}\frac{(\vec{S}_i\cdot\vec{S}_k)(\vec{S}_k\cdot\vec{S}_j)
    +(\vec{S}_j\cdot\vec{S}_k)(\vec{S}_k\cdot\vec{S}_i)}{2(S+1)(2S+3)(\bar{z}_k-\bar{z}_i)(z_k-z_j)}
    +\sum_k\sum_{i,j(\neq k)}\frac{(\vec{S}_i\cdot\vec{S}_k)(\vec{S}_k\cdot\vec{S}_j)}{4(S+1)(2S+3)}\Biggr\}\;.
\end{align}
  We see that the \iMPS Hamiltonian features Haldane-Shastry-like spin-spin interactions as well as three-spin interactions, all of them long-ranged. In addition there is a term proportional to $\vec{S}_{\text{tot}}^2$ that can be interpreted as some sort of chemical potential that is favoring large total spins. It is likely though that this term will be dominated by contributions arising from the final three-spin term and that, overall, small spins will be favoured. When carrying out the sums over $i$ and $j$, the final term can be written in terms of the total spin $\vec{S}_{\text{tot}}$ but is is unclear how to simplify the sum $\sum_k(\vec{S}_k\cdot\vec{S}_{\text{tot}})^2$ that arises in this way.

  A closer inspection reveals that the Hamiltonian \eqref{eq:HFinal} is similar but not quite identical to the one that was derived in Ref.~\cite{Greiter:2011jp}. Indeed, while the $z$-dependent (and hence distance dependent) interaction terms precisely agree, the Hamiltonian~\eqref{eq:HFinal} features additional terms which have the flavor of chemical potentials. From a naive perspective it looks possible to remedy the issue by working with one of the other two choices for $\Omega_{ij}$ that we have mentioned at the beginning of this section, namely $x_{ij}$ or $\theta_{ij}$. Indeed, these different choices are related by
\begin{align}
  \label{eq:Change}
  |w_{ij}|^2+1
  &=4|\theta_{ij}|^2
   =4|x_{ij}|^2
   =-\frac{4z_iz_j}{(z_i-z_j)^2}
   =\frac{4}{|z_i-z_j|^2}\nonumber\\[2mm]
  \bar{w}_{ki}w_{kj}
  &=2(\bar{\theta}_{ki}\theta_{kj}+\bar{\theta}_{kj}\theta_{kj})-1
   =2(\bar{x}_{ki}x_{kj}+\bar{x}_{kj}x_{kj})-1\;,
\end{align}
  so they are guaranteed to give the desired dependence on the parameters $z_l$ when inserted into the Hamiltonian~\eqref{eq:HSymmetrized}. On the other hand, for both of these choices the three-spin term $(\vec{S}_i\times\vec{S}_k)\cdot\vec{S}_j$ does not drop out, which results in a Hamiltonian that is manifestly chiral. While this may be desired in a 2D setting, it seems rather unnatural in one dimension, at least if one wishes to find a realization of the $SU(2)_k$ WZW model.


\section{Conclusions}

  In this paper we have applied the \iMPS construction to $SU(2)_k$ WZW models. Using a basis-independent formalism for spin operators and projectors allowed us to derive explicit parent Hamiltonians for 1D and 2D quantum spin models. We also established a precise correspondence between our conformal field theory approach and the symmetrization approach for trial wavefunctions of a family of non-abelian chiral spin liquids \cite{Greiter:2009PhRvL.102t7203G}. Even though the wavefunctions agree there seems to be a slight mismatch in the associated parent Hamiltonians. To our knowledge, this paper has been the first complete and explicit application of the \iMPS construction to a family of truly interacting conformal field theories after earlier papers on $U(1)$, $SU(2)_1$, $SO(N)_1$ and $SU(N)_1$ WZW models \cite{Cirac:2010PhRvB..81j4431C,Nielsen:2011py,Tu:2013PhRvB..87d1103T,Tu:2014NuPhB.886..328T,Bondesan:2014NuPhB.886..483B} which can all be interpreted as free field theories (with the special case $SO(3)_1\cong SU(2)_2$).

  Our present discussion focused on 2D quantum spin systems on the plane (or Riemann sphere) and closed 1D systems on the circle. It is known that the \iMPS construction generalizes to 1D systems with open boundary conditions \cite{Tu:2015PhRvB..92d1119T,BasuMallick:2016PhRvB..93o5154B} and to systems on the torus \cite{Nielsen:2014JSMTE..04..007N,Deshpande:2016JSMTE..01.3102D} and it would be interesting to carry these constructions out for $SU(2)_k$. Moreover, the non-abelian chiral spin liquid we constructed was obtained by combining $k$ {\em identical} copies of the state for $k=1$. It would be worthwhile to understand how the more general possibilities of combining $k=1$ theories as discussed in Ref.~\cite{Scharfenberger:2011PhRvB..84n0404S} can be realized in the framework of \iMPS.
  
  Finally, the last few years have seen tremendous efforts to understand abelian and non-abelian spin liquids using what is called a coupled wire construction \cite{Gorohovsky:2015PhRvB..91x5139G,Meng:2015PhRvB..91x1106M,Huang:2016PhRvB..93t5123H} and generalized spin ladders \cite{Lecheminant:2017PhRvB..95n0406L}. While there are certain similarities to the approach discussed in the present paper it still needs to be verified whether these are merely superficial or whether both constructions have a more intimate relationship.

  

\subsubsection*{Acknowledgements}

  AR and TQ gratefully acknowlegde discussions with Roberto Bondesan and collaboration on closely related topics.
  This research was conducted by the Australian Research Council Centre of Excellence for Mathematical and Statistical Frontiers (project number CE140100049) and partially funded by the Australian Government.
  Parts of this work were carried out while the authors were employed at the Institute of Theoretical Physics at the University of Cologne. We would like to thank the DFG for financial support through M.\ Zirnbauer’s Leibniz Prize, DFG grant no. ZI 513/2-1. Additional support was received from the DFG through the SFB$|$TR 12 ``Symmetries and Universality in
Mesoscopic Systems'' and the Center of Excellence ``Quantum Matter and Materials''.

\appendix
\section{\label{ap:Notation}Summary of notation and conventions}

  Throughout the paper we will work with a basis of spin operators $S^a$ (with $a=1,2,3$) that satisfy
\begin{align}
  [S^a,S^b]
  =i{\epsilon^{ab}}_c\,S^c\;,
\end{align}
  where ${\epsilon^{ab}}_c$ is completely antisymmetric in its indices and ${\epsilon^{12}}_3=1$. Indices are raised (and lowered) with the Killing form $\kappa^{ab}=\delta^{ab}$ (and its inverse $\kappa_{ab}=\delta_{ab}$). For certain simplifications we will need the identities
\begin{align}
  \label{eq:epsidentity}
  \epsilon^{abc}\,\epsilon_{abd}
  =2\,\delta^c_d
  \quad\text{ and }\quad
  \epsilon^{abe}{\epsilon^{cd}}_e
  =\delta^{ac}\delta^{bd}-\delta^{ad}\delta^{bc}\;.
\end{align}
  In writing these (and other) formulas Einstein's summation convention is understood, i.e.\ indices occurring twice on one side of an equation are summed over the relevant range.

\section{\label{ap:SchwingerBosons}Schwinger bosons}

  The essential idea of the Schwinger boson construction is to realize all finite dimensional irreducible representations of $SU(2)$ on the Fock space of two bosons~\cite{SchwingerBosons}. As we will review shortly, individual representations can be singled out simply by projection onto subspaces of specific boson number.
  
  Let $a^\dag,a$ and $b^\dag,b$ be the creation and annihilation operators of the two bosons. By definition, these satisfy
\begin{align}
  [a,a^\dag]=[b,b^\dag]=1\;,
\end{align}
  with all other commutators of these generators vanishing.
  From these we define spin operators
\begin{align}
  S^z
  &=\frac{1}{2}(a^\dag a-b^\dag b)\;,&
  S^+
  &=a^\dag b\;,&
  S^-
  &=b^\dag a
\end{align}
  that satisfy the desired commutation relations of $SU(2)$.
  The Fock space of the Schwinger bosons hence realizes a representation of $SU(2)$, containing the irreducible representations $S=0,\frac{1}{2},1,\ldots$. In view of the relation
\begin{align}
  \vec{S}^2
  =\frac{1}{4}\,(a^\dag a+b^\dag b)(a^\dag a+b^\dag b+2)
  =S(S+1)
\end{align}
  half the total boson number
\begin{align}
  S=\frac{1}{2}\bigl(aa^\dag+bb^\dag\bigr)
\end{align}
  may be interpreted as the spin of the representation that is realized.
  
  The Fock space is created from a vacuum vector $|0\rangle$ defined by $a|0\rangle=b|0\rangle=0$. An orthonormal basis of states is obtained by defining
\begin{align}
  |S,m\rangle
  =\frac{(a^\dag)^{S+m}}{\sqrt{(S+m)!}}
   \frac{(b^\dag)^{S-m}}{\sqrt{(S-m)!}}\,|0\rangle\;.
\end{align}
  We finally note that
\begin{align}
  |S,-S\rangle
  =\frac{(b^\dag)^{2S}}{\sqrt{(2S)!}}\,|0\rangle\;.
\end{align}
  This is the state in the spin-$S$ representation with minimal value of $S^z$ (spin maximally down).
  The Schwinger bosons may be used to represent the two states of the spin-$\frac{1}{2}$ representation as follows,
\begin{align}
  |\!\uparrow\rangle
  =a^\dag|0\rangle
  \quad\text{ and }\quad
  |\!\downarrow\rangle
  =b^\dag|0\rangle\,
\end{align}
  and similarly for the three states of the spin-$1$ representation,
\begin{align}
  |\!\Uparrow\rangle
  =\frac{1}{\sqrt{2}}(a^\dag)^2|0\rangle\;,\quad
  |\!\Leftrightarrow\rangle
  =a^\dag b^\dag|0\rangle
  \quad\text{ and }\quad
  |\!\Downarrow\rangle
  =\frac{1}{\sqrt{2}}(b^\dag)^2|0\rangle\;.
\end{align}

\section{\label{ap:GeneralG}Simple current correlators for other symmetry groups}

\begin{table}
\begin{center}
\begin{tabular}{cccc}
  Type & Group & Simple current(s) & Conformal dimension \\\hline\hline\\[-1em]
  $A$ & $SU(N)$ & $j_l=k\omega_l$ & $h_{j_l}=\frac{kl(N-l)}{2N}$ \\[2mm]
  $B$ & $SO(2N+1)$ & $j=k\omega_1$ & $h_j=\frac{k}{2}$ \\[2mm]
  $C$ & $SP(2N)$ & $j=k\omega_N$ & $h_j=\frac{kN}{4}$ \\[2mm]
  $D$ & $SO(2N)$ & $j_1=k\omega_1$ & $h_{j_1}=\frac{k}{2}$ \\[2mm]
    && $j_2=k\omega_{N-1}$ & $h_{j_2}=\frac{kN}{8}$\\[2mm]
    && $j_3=k\omega_N$ & $h_{j_3}=\frac{kN}{8}$\\[2mm]
  $E$ & $E_6$ & $j_1=k\omega_1$ & $h_{j_1}=\frac{2k}{3}$ \\[2mm]
    && $j_2=k\omega_5$ & $h_{j_2}=\frac{2k}{3}$ \\[2mm]
  $E$ & $E_7$ & $j=k\omega_6$ & $h_j=\frac{3k}{4}$
\end{tabular}
  \caption{\label{tab:SimpleCurrents}Simple currents in generic level-$k$ WZW models and their conformal dimensions.}
\end{center}
\end{table}

  The considerations for the ground state wave function presented in Section~\ref{sc:GroundState} carry over to other WZW models at generic level $k$ based on simple Lie groups. This is essentially due to the relation $h_{j_k}=kh_{j_1}$ where $j_k$ is a simple current in the level-$k$ theory that arises from the symmetric combination of $k$ simple currents $j_1$ in $k$ distinct level-$1$ theories based on the same group. The relevant relation may be verified case by case and a more detailed discussion can be found in the book~\cite{Fuchs:1995}, see especially the paragraphs around Eqs.~(3.5.43) and~(5.1.30). Here we content ourselves with the following table that summarizes the simple currents at generic levels (in the labeling conventions of \cite{FrancescoCFT}), together with their conformal dimensions.


\def\cprime{$'$}
\providecommand{\href}[2]{#2}\begingroup\raggedright\endgroup

\end{document}